\documentclass[twocolumn,showpacs,aps,prl,superscriptaddress,preprintnumbers,amsmath,amssymb,floatfix]{revtex4}
\usepackage{graphicx}% Include figure files
\usepackage{dcolumn}% Align table columns on decimal point
\usepackage{bm}% bold math

\usepackage{relsize}
\usepackage{xspace}
\def\babar{\mbox{\slshape B\kern-0.1em{\smaller A}\kern-0.1em
    B\kern-0.1em{\smaller A\kern-0.2em R}}}
\def\pep2{PEP-II}
\def\D0bar{\kern 0.2em\overline{\kern -0.2em D}{\kern 0.1em}\xspace^0}

\newcommand{\SLACPubNumber} {9888}

\newcommand{\gevc}{\ensuremath{{\mathrm{\,Ge\kern -0.1em V\!/}c}}\xspace}
\newcommand{\mevc}{\ensuremath{{\mathrm{\,Me\kern -0.1em V\!/}c}}\xspace}
\newcommand{\gevcc}{\ensuremath{{\mathrm{\,Ge\kern -0.1em V\!/}c^2}}\xspace}
\newcommand{\mevcc}{\ensuremath{{\mathrm{\,Me\kern -0.1em V\!/}c^2}}\xspace}

\def\CP{\ensuremath{C\!P}\xspace}

\begin{document}

%\preprint{\babar\  Analysis Document \#587, Version 8}
%\preprint{BABAR-PUB-\BABARPubYear/\BABARPubNumber}
\preprint{SLAC-PUB-\SLACPubNumber}
\preprint{hep-ex/0306003}

\title{\boldmath Limits on $D^0$-$\D0bar$ Mixing and \CP Violation
from the Ratio of Lifetimes for Decay to $K^-\pi^+$, $K^-K^+$, and $\pi^-\pi^+$}

%% author list as of 02-Apr-2003 (577 authors)
%
\author{B.~Aubert}
\author{R.~Barate}
\author{D.~Boutigny}
\author{J.-M.~Gaillard}
\author{A.~Hicheur}
\author{Y.~Karyotakis}
\author{J.~P.~Lees}
\author{P.~Robbe}
\author{V.~Tisserand}
\author{A.~Zghiche}
\affiliation{Laboratoire de Physique des Particules, F-74941 Annecy-le-Vieux, France }
% ...
\collaboration{The \babar\ Collaboration}
\noaffiliation

\date{\today}% It is always \today, today,

\begin{abstract}
We present a measurement of $D^0$-$\D0bar$ mixing parameters
using the ratios of lifetimes extracted from samples of $D^0$ mesons
decaying to $K^-\pi^+$, $K^-K^+$, and $\pi^-\pi^+$.
Using 91~${\rm fb}^{-1}$ of
data collected by the \babar\  detector at the PEP-II
asymmetric-energy $B$ Factory, we obtain a value
$Y = (0.8 \pm 0.4 ({\rm stat.}) \mbox{}^{+0.5}_{-0.4} ({\rm syst.}))$\%,
which, in the limit of \CP conservation, corresponds to the
mixing parameter $y = \Delta\Gamma/2\Gamma$. Using the difference in lifetimes
of $D^0$ and $\D0bar$ mesons, we obtain the \CP-violation parameter 
$\Delta Y  = (-0.8 \pm 0.6 ({\rm stat.}) \pm 0.2 ({\rm syst.}))$\%.
\end{abstract}

\maketitle

To date there is no experimental evidence for mixing in the $D^0$-$\D0bar$
system~\cite{Hagiwara:2002pw,Grothe:2003zg}. 
This is consistent with Standard Model 
expectations~\cite{Falk:2001hx,Nelson:1999fg}, which correspond to
a level of mixing beyond the reach of current
experimental precision. 
Among the more striking consequences of
$D^0$-$\D0bar$ mixing are different decay-time distributions for
$D^0$ mesons that decay into final states of specific 
\CP~\cite{Liu:1994ea}.
Measurable \CP violation in $D^0$-$\D0bar$ mixing would be 
evidence of physics beyond the Standard Model~\cite{Blaylock:1995ay}.

The two $D^0$ mass eigenstates can be represented as
\begin{equation}
\begin{array}{rcl}
| D_1 \rangle &=& p | D^0 \rangle + q | \D0bar \rangle \\
| D_2 \rangle &=& p | D^0 \rangle - q | \D0bar \rangle \;,
\end{array}
\label{eq:qpdef}
\end{equation}
where $\left|p\right|^2 + \left|q\right|^2 = 1$.
It is traditional to quantify the size of $D^0$-$\D0bar$ mixing
in terms of the parameters $x \equiv \Delta m/\Gamma$ and
$y \equiv \Delta\Gamma/2\Gamma$, where $\Delta m = m_1 - m_2$
($\Delta \Gamma = \Gamma_1 - \Gamma_2$) is the difference in
mass (width) of the states of Eq.~(\ref{eq:qpdef})
and $\Gamma = (\Gamma_1+\Gamma_2)/2$ is the average width.
If either $x$ or $y$ is non-zero, mixing will occur.
The Standard Model expectation for the size of both is 
$\lesssim 10^{-3}$~\cite{Falk:2001hx,Nelson:1999fg}.

The effects of \CP violation in $D^0$-$\D0bar$ mixing
can be parameterized in terms of the quantities
\begin{equation}
r_m \equiv \left| \frac{q}{p} \right| 
\mbox{\hskip 0.25in}{\rm and}\mbox{\hskip 0.25in}
\varphi_f \equiv \arg\left( \frac{q}{p}\frac{\overline{A}_f}{A_f} \right) \;,
\end{equation}
where $A_f \equiv \langle f | {\mathcal H}_D | D^0 \rangle$ 
($\overline{A}_f \equiv \langle f | {\mathcal H}_D | \D0bar \rangle$) is the
amplitude for $D^0$ ($\D0bar$) decaying into a final state
$f$. A value of $r_m \neq 1$ would indicate \CP violation in mixing.
A non-zero value of $\varphi_f$ would indicate \CP violation
in the interference of mixing and decay. Direct 
\CP violation
is expected to be small in the $D^0$-$\D0bar$ system~\cite{Bergmann:2000id}
and is not considered here.

$D^0$-$\D0bar$ mixing will alter the decay time distribution
of $D^0$ and $\D0bar$ mesons that decay into
final states of specific \CP. 
To a good approximation, these decay
time distributions can be treated as exponential
with effective lifetimes~\cite{Bergmann:2000id}
\begin{equation}
\begin{array}{rcl}
\tau^+   &=& 
\tau^0\left[ 1 + r_m \left( y\cos\varphi_f - 
                                 x\sin\varphi_f \right) \right]^{-1} \\
\tau^- &=&
\tau^0\left[ 1 + r_m^{-1} \left( y\cos\varphi_f +
                                      x\sin\varphi_f \right) \right]^{-1} \;,
\end{array}
\end{equation}
where $\tau^0$ is the lifetime for the Cabibbo-favored decays
$D^0 \to K^-\pi^+$ and $\D0bar \to K^+\pi^-$ and $\tau^+$ ($\tau^-$)
is the lifetime for the Cabibbo-suppressed decays
of the $D^0$ ($\D0bar$) into \CP-even final states (such as $K^-K^+$
and $\pi^-\pi^+$). 
These effective lifetimes can be combined into the 
following quantities $Y$ and $\Delta Y$:
\begin{equation}
Y = \frac{\tau^0}{\langle\tau\rangle} - 1
\mbox{\hskip 0.5in}
\Delta Y = \frac{\tau^0}{\langle\tau\rangle} A_\tau \;,
\end{equation}
where $\langle \tau \rangle = (\tau^++\tau^-)/2$ and
$A_\tau = (\tau^+-\tau^-)/(\tau^++\tau^-)$. Both $Y$ and $\Delta Y$
are zero if there is no $D^0$-$\D0bar$ mixing. Otherwise,
in the limit of \CP conservation in mixing,
$Y = y\cos\varphi_f$ and $\Delta Y = x\sin\varphi_f$.

We present a measurement of $Y$ and $\Delta Y$ 
obtained from
a 91~${\rm fb}^{-1}$
data sample collected on or near the $\Upsilon(4S)$ resonance
with the \babar\  detector at the
\pep2 asymmetric-energy $e^+e^-$ storage ring.

The \babar\ detector,
a general-purpose, solenoidal, magnetic spectrometer, is
described in more detail elsewhere \cite{Aubert:2001tu}. 
Charged particles were detected
and their momenta measured by a combination of a drift chamber (DCH)
and silicon vertex tracker (SVT), both operating within a
1.5-T solenoidal magnetic field. 
A ring-imaging Cherenkov detector (DIRC) was used for
charged-particle identification.

Four independent samples of $D^0$ and $\D0bar$
mesons were used in this analysis.
The first three samples (referred to as {\it tagged})
correspond to $D^0$ mesons that 
decayed into $K^-\pi^+$, $K^-K^+$, and $\pi^-\pi^+$~\footnote{
Unless otherwise noted, statements involving $D^0$ mesons
and their decay modes are intended to apply in addition to their
charged conjugates.} and
include the decay $D^{*+} \rightarrow D^0 \pi^+$
to suppress backgrounds and distinguish $D^0$ from $\D0bar$. 
These three samples were used to measure $Y$ and 
$\Delta Y$.
The fourth sample (referred to as {\it untagged}) consisted of $K^-K^+$ decays
that were not matched to a $D^{*+}$ decay and was used 
to measure $Y$.

$D^0$ candidates were selected by searching for pairs of 
oppositely charged tracks of invariant mass near the 
expected value for a $D^0$ meson. Each track
was required to contain a minimum number of measurement points in the
SVT and DCH. The two $D^0$-candidate daughter
tracks were fitted to
a common vertex. The fit probability of this vertex fit was
required to be larger than 1\%. The interaction point (IP)
was determined by calculating the point in space most
consistent with the $D^0$ trajectory and the beam envelope
(approximately 6~$\mu$m high and 120~$\mu$m wide).

Each $D^0$ daughter track was subjected to a likelihood-based particle
identification algorithm. This algorithm relied on the measurement
of the Cherenkov angle from the DIRC and on the energy loss ($dE/dx$) 
measured with the SVT and DCH.
The $K^\pm$ identification efficiency was approximately
80\% 
for tracks within the DIRC acceptance with a $\pi^\pm$ misidentification
probability of about 2\%. The average $\pi^\pm$ identification efficiency
was approximately 90\%.

To reduce combinatorial background that tended to accumulate at lower momenta,
each $D^0$ candidate was required to have a momentum in the $e^+e^-$
center-of-mass
frame greater than 2.4~\gevc. This requirement was also effective at
removing $D^0$ mesons originating from the decays of $B$ mesons.

The proper decay time and its measurement error $\sigma_t$
for each $D^0$ candidate were calculated using the $D^0$ and IP
vertex fits. The world average $D^0$ mass~\cite{Hagiwara:2002pw} $m_D$
and the  momentum of the $D^0$ were used to calculate the boost of the $D^0$
and to obtain the proper decay time. The distribution of $\sigma_t$,
uncorrelated with true decay time, peaks at
a value of 160~fs, and has a long upper tail.
Poorly measured $D^0$ candidates 
with $\sigma_t > 500$~fs (16\% of each sample) were discarded.

The decay $D^{*+} \rightarrow D^0 \pi^+$ is characterized by a
$\pi^+$ of low momentum ($\pi_s$). 
To increase acceptance, $\pi_s$ candidate tracks were not
required to include DCH measurements. To improve momentum
resolution, a vertex fit
was used to constrain each $\pi_s$
candidate track to pass through the IP. If the fit
probability of this
vertex fit was less than 1\%, the $D^{*+}$ candidate was discarded.

The distribution of the difference
in the reconstructed $D^{*+}$ and $D^0$ masses ($\delta m$)
peaked near $145.4$~\mevcc. Backgrounds were suppressed
by discarding $D^{*+}$ candidates with a value of $\delta m$ that deviated 
more than 1 (2.5)~\mevcc from the peak for those $\pi_s$ tracks 
measured with (without) the DCH.

\begin{figure}
\includegraphics[width=\linewidth]{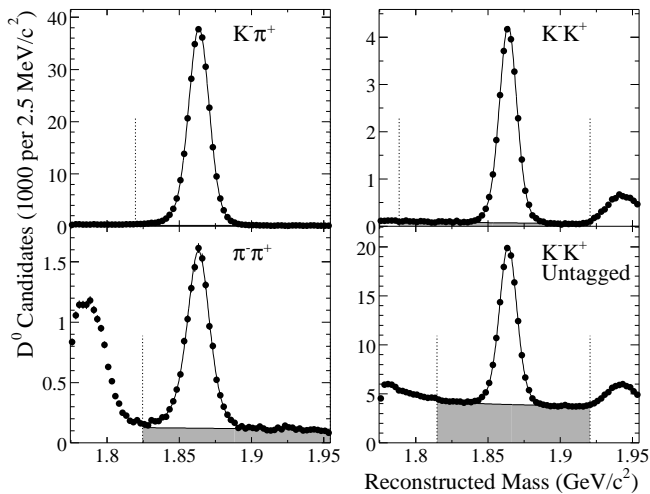}
\caption{\label{fg:PrlMassFit}The reconstructed $D^0$ mass distribution 
(points) superimposed on a projection of the mass fit (curve)
for the four $D^0$ samples.
The fit was performed within the restricted ranges of mass indicated
by the vertical dotted lines.
The portion of the sample assigned by the fit to the background
is indicated by the shaded region.
}
\end{figure}

The $D^0$ mass 
distributions for the selected $D^0$ candidates are shown in
Fig.~\ref{fg:PrlMassFit}.
Ample sidebands were included to
measure the characteristics of the background. The peaks appearing
above or below the $D^0$ mass were due to candidates
with misidentified kaons or pions. 
For presentation purposes
only, we define those $D^0$ candidates with reconstructed
masses within 15~\mevcc of $m_D$ as belonging to a mass signal window.
The sizes and estimated purities of the four $D^0$ samples
within this window are listed in Table~\ref{tb:sample}.

\begin{table}
\caption{\label{tb:sample}The four $D^0$ samples, their use, and,
as calculated inside a $\pm 15$~\mevcc mass window, their
size, and purity.}
\begin{ruledtabular}
\begin{tabular}{lcrr}
Sample & Measures & Size & Purity (\%) \\
\hline
$K^-\pi^+$         & $\tau^0$                           & 265,152 & 99.4  \\
$K^-K^+$           & $\langle \tau \rangle$, $A_{\tau}$ &  26,084 & 97.0  \\
$\pi^-\pi^+$       & $\langle \tau \rangle$, $A_{\tau}$ &  12,849 & 87.9  \\
Untagged $K^-K^+$  & $\langle \tau \rangle$             & 145,826 & 68.1  \\
\end{tabular}
\end{ruledtabular}
\end{table}

An unbinned maximum-likelihood fit was used to extract the lifetime
from each $D^0$ sample. 
The likelihood function consisted of two decay-time distribution
functions, one for signal and one for background. The signal
function was a convolution of an exponential and
a resolution function that was the sum of
three Gaussian distributions with zero mean.
The widths of the first two Gaussians were
proportional to $\sigma_t$ whereas the width of the third,
designed to describe mismeasurements, was not.
The parameters in the fit associated with the signal 
for the $K^-\pi^+$ and untagged $K^-K^+$ samples were
the lifetime and the widths and relative proportions of the three Gaussians.
The parameters for the tagged $K^-K^+$ and $\pi^-\pi^+$ samples
were the same except for the addition of $A_\tau$.

As in the signal likelihood function, 
the background function was a convolution of
a resolution function and a lifetime distribution.
The background lifetime distribution was the sum of an
exponential distribution and a delta function at zero, the
latter corresponding to prompt sources of 
background that originated at the IP.
The resolution function consisted of the sum
of four Gaussian distributions, the first three of which
were similar to those of the signal. The fourth was given a fixed
width of 12~ps and accounted for a small number ($<10^{-3}$) of outliers
produced by long-lived particles or reconstruction errors.
The additional fit parameters associated with the background included the
fraction assigned to zero lifetime sources, 
the background lifetime, and the relative size of the fourth
Gaussian.

To combine the signal and background likelihood functions,
the reconstructed mass of each $D^0$ candidate was used to
determine the probability that it was a signal $D^0$.
This calculation was based on a separate fit of the
reconstructed $D^0$ mass distribution (Fig.~\ref{fg:PrlMassFit}).
This fit included a resolution function composed of 
a Gaussian with an asymmetric tail designed to account for final-state
photon radiation. The mass fit for the tagged $D^0$ samples included
a linear portion to describe the
background. The slope of the background was constrained with
$D^0$ candidates in the $\delta m$ sideband ($151 < \delta m < 159$~\mevcc).
For the untagged $K^-K^+$ sample, the size of the radiative tail was 
taken from the tagged $K^-K^+$ sample and the background was described by a
quadratic function.

\begin{figure}
\includegraphics[width=\linewidth]{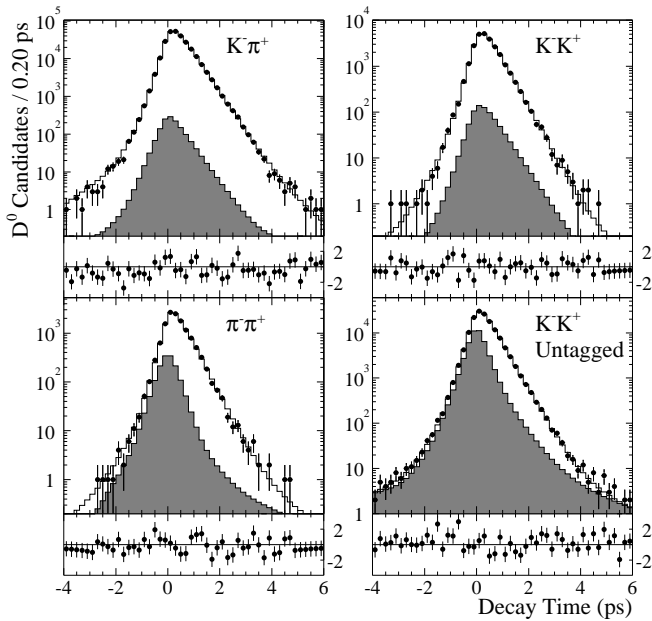}
\caption{\label{fg:PrlLife} The decay-time distribution for the
four $D^0$ samples (points) within a $\pm 15$~\mevcc mass signal window 
superimposed on a projection of the lifetime fit (histogram).
The shaded histogram is the portion of the sample assigned by the fit to
the background. The points presented below the histograms are the
difference between data and fit divided by the statistical error with
error bars of unit length.}
\end{figure}

The results of the lifetime fits are shown in Fig.~\ref{fg:PrlLife}.
Typical values for the fit parameters were 
a background lifetime similar to the $D^0$ lifetime
and a third Gaussian width that was several times larger than
the typical decay-time error. 
%The proportionality factors associated with the 
%first two Gaussians in the signal resolution function corresponded to
%a root-mean-square of approximately $1.1$.

To ensure that the analysis was performed in an objective manner,
the values of the $\tau^0$, $\langle\tau\rangle$, and $A_\tau$ 
fit parameters were hidden until
the analysis method and systematic uncertainties were finalized.

Potential biases in $Y$ and $\Delta Y$ were investigated using
Monte Carlo (MC) samples produced by a GEANT4-based~\cite{Agostinelli:2002hh} 
detector simulation
and processed by the same reconstruction and analysis programs as the
data. To estimate the behavior of both signal and background,
the equivalent of 70~${\rm fb}^{-1}$ of continuum and $\Upsilon(4S)$
MC data were studied. To augment these samples, separate MC
samples were generated for specific decay modes of the $D^0$ meson. 
The values of $Y$ and
$\Delta Y$ calculated from the MC samples (generated with $Y=0$ 
and $\Delta Y=0$) were consistent with zero within
statistical errors. 
As can be seen in Fig.~\ref{fg:PrlLife}, the fraction of prompt 
background is larger in the $\pi^-\pi^+$ and untagged $K^-K^+$ samples
than in the other two, a tendency that is accurately reproduced by the
MC simulation.

Potential inaccuracies in the simulation of tracking were explored
by varying within current understanding 
the assumptions used in the MC simulation, including the center-of-mass
boost and energy, the strength of the magnetic field, and tracking
resolution and efficiency. Small charge asymmetries 
(at the level of 2 to 4\%) in the reconstruction efficiency
of $\pi_s$ tracks produced slightly different momentum and angular
distributions for $D^0$ and $\D0bar$ mesons. The influence of
these effects on $Y$ and $\Delta Y$ was checked by applying the lifetime
fits on the data after weighting events to remove these asymmetries.

The size and decay time characteristics of backgrounds in the data 
were determined in the likelihood fits without using any MC input. The MC 
samples were used, however, to 
determine how well the fits account for the presence
of background. As part of these tests, the size of specific types of
backgrounds was varied within uncertainties 
in the MC sample by reweighting. In addition,
MC parameters associated with the charm fragmentation function
and final-state radiation in $D^0$ decay were varied.
The resulting
effect on the fitted lifetime is reported as one source of systematic
uncertainty.

Detector misalignment was another potential source of bias.
Residual distortions of the SVT,
even as small as a few microns, can produce significant
variations in the apparent $D^0$ lifetime. 
Several studies were
used to measure and characterize such distortions and
strategies were developed to correct them. 
One example was the study of
proton tracks that were created by the interaction of off-energy
beam particles and the beampipe. These tracks were used to
measure the radius of the beampipe to a precision of a few microns,
which limited the uncertainty in the radial scale of
the SVT to 0.3\%.

Another example was a study of
$e^+ e^-  \rightarrow e^+ e^- + 2(\pi^+\pi^-)$ 
events in which the four pions were known to originate from the IP.
By selecting oppositely charged pairs of these pions with opening
angles similar to two-body $D^0$ decays, it was possible to
measure the apparent beam position as a function of $D^0$-candidate
trajectory
and calculate a correction to the $D^0$ lifetime. 
This type of correction nearly
cancels in the lifetime ratio and introduces little systematic uncertainty
in $Y$ or $\Delta Y$.

Because $Y$ and $\Delta Y$ were measured from the ratio and asymmetry
of lifetimes, systematic uncertainties from alignment that have
a strong influence on $\tau^0$ did not make a large contribution
to $Y$ and $\Delta Y$.
Efforts to reduce these systematic uncertainties are
still underway; therefore, a 
value of $\tau^0$ is not reported in this letter.
A subsample of the data was used to verify that $\tau^0$ is
consistent within uncertainties with the world average~\cite{Hagiwara:2002pw}.

%    Suggestion from Brian:
% These data are clearly sufficient to improve on the current PDG
% average for the D0 lifetime, though systematic effects are significant.
% The BaBar alignment, though adequate for the present study in which
% its effects largely cancel, can be further improved.  We therefore
% choose to keep the actual lifetime value blinded until all
% improvement attempts have been exhausted.
 
The systematic uncertainties in $Y$ and $\Delta Y$ are
summarized in Table~\ref{tb:PrlRatioSyst}. The separate results
for each sample are listed in
Table~\ref{tb:PrlResults} with combined values that assume
the same value of $\varphi_f$ for the $K^-K^+$ and $\pi^-\pi^+$ decay modes.
All values are consistent with no mixing.
Because it is derived from an asymmetry, 
the systematic uncertainty in $\Delta Y$ is considerably smaller than in $Y$.

\begin{table}
\caption{\label{tb:PrlRatioSyst}Summary of systematic uncertainties.}
\renewcommand{\arraystretch}{1.2}
\begin{ruledtabular}
%autogenerated: Mon May 19 14:15:03 2003
\begin{tabular}
{lrrrrr}
\multicolumn{1}{c}{} &
\multicolumn{3}{c}{Change in $Y$ (\%)} &
\multicolumn{2}{c}{$\Delta Y$ (\%)} \\
\cline{2-4}
\cline{5-6}
\multicolumn{1}{c}{} &
\multicolumn{2}{c}{Tagged} &
\multicolumn{1}{c}{Untag} &
\multicolumn{2}{c}{Tagged} \\
\cline{2-3}
\cline{4-4}
\cline{5-6}
Category &
$K^-K^+$ &
$\pi^-\pi^+$ &
$K^-K^+$ &
$K^-K^+$ &
$\pi^-\pi^+$ \\
\hline
Tracking
  & $\pm\,0.1$
  & $\pm\,0.3$
  & $\pm\,0.2$
  & $\pm\,0.1$
  & $\pm\,0.1$
  \\
Background
  & $ \mbox{}^{ +\,0.3 }_{ -\,0.5 } $ 
  & $\pm\,0.5$
  & $\pm\,0.3$
  & $\pm\,0.2$
  & $\pm\,0.2$
  \\
Alignment
  & $\pm\,0.1$
  & $\pm\,0.1$
  & $\pm\,0.1$
  & $\pm\,0.1$
  & $ \mbox{}^{ +\,0.0 }_{ -\,0.1 } $ 
  \\
MC Statistics
  & $ \mbox{}^{ +\,0.4 }_{ -\,0.1 } $ 
  & $ \mbox{}^{ +\,1.0 }_{ -\,0.1 } $ 
  & $ \mbox{}^{ +\,0.4 }_{ -\,0.1 } $ 
  & $<0.1$
  & $<0.1$
  \\
\hline
Quadrature Sum
  & $\pm\,0.5$
  & $ \mbox{}^{ +\,1.2 }_{ -\,0.6 } $ 
  & $ \mbox{}^{ +\,0.5 }_{ -\,0.4 } $ 
  & $\pm\,0.2$
  & $\pm\,0.2$
  \\
\end{tabular}

\end{ruledtabular}
\end{table}

\begin{table}
\caption{\label{tb:PrlResults}Summary of $Y$ and $\Delta Y$ results.
The first error is statistical; the second, systematic.}
\renewcommand{\arraystretch}{1.2}
\begin{ruledtabular}
%autogenerated: Tue Apr  8 11:47:35 2003
\begin{tabular}{l@{\hskip 0.35in}r@{.}l@{}c@{}c@{\,}c@{\hskip 0.35in}r@{.}l@{}c@{}c@{\,}c}
Sample & \multicolumn{5}{c@{\hskip 0.35in}}{$Y$ (\%)}
       & \multicolumn{5}{c}{$\Delta Y$ (\%)} \\
\hline
$K^-K^+$
 & $1$&$5$
 &$\:\pm\:$ 
 & $0.8$
 & $\pm\,0.5$
 & $-1$&$3$
 &$\:\pm\:$ 
 & $0.8$
 & $\pm\,0.2$
\\
$\pi^-\pi^+$
 & $1$&$7$
 &$\:\pm\:$ 
 & $1.2$
 & $ \mbox{}^{ +\,1.2 }_{ -\,0.6 } $ 
 & $0$&$3$
 &$\:\pm\:$ 
 & $1.1$
 & $\pm\,0.2$
\\
Untagged $K^-K^+$
 & $0$&$2$
 &$\:\pm\:$ 
 & $0.5$
 & $ \mbox{}^{ +\,0.5 }_{ -\,0.4 } $ 
 &\multicolumn{5}{c}{---}\\
\hline
Combined
 & $0$&$8$
 &$\:\pm\:$ 
 & $0.4$
 & $ \mbox{}^{ +\,0.5 }_{ -\,0.4 } $ 
 & $-0$&$8$
 &$\:\pm\:$ 
 & $0.6$
 & $\pm\,0.2$
\\
\end{tabular}

\end{ruledtabular}
\end{table}

In summary, we have obtained a value of 
$Y = (0.8 \pm 0.4 ({\rm stat.}) \mbox{}^{+0.5}_{-0.4} ({\rm syst.}))$\% that 
is consistent
with no mixing and is at least twice as precise as
previous measurements of this 
type~\cite{Abe:2001ed,Cronin-Hennessy:2001cw,Aitala:1999dt,Link:2000cu}, 
all of which assumed \CP conservation.
We also obtain for the first time a measurement of 
$\Delta Y = (-0.8 \pm 0.6 ({\rm stat.}) \pm 0.2 ({\rm syst.}))$\%.

\begin{acknowledgments}
We are grateful for the excellent luminosity and machine conditions
provided by our \pep2\ colleagues, 
and for the substantial dedicated effort from
the computing organizations that support \babar.
The collaborating institutions wish to thank 
SLAC for its support and kind hospitality. 
This work is supported by
DOE
and NSF (USA),
NSERC (Canada),
IHEP (China),
CEA and
CNRS-IN2P3
(France),
BMBF and DFG
(Germany),
INFN (Italy),
FOM (The Netherlands),
NFR (Norway),
MIST (Russia), and
PPARC (United Kingdom). 
Individuals have received support from the 
A.~P.~Sloan Foundation, 
Research Corporation,
and Alexander von Humboldt Foundation.

\end{acknowledgments}

%\newpage %Just because of unusual number of tables stacked at end
\bibliography{mixing}% Produces the bibliography via BibTeX.

% 55 lines for an author list
%54\\53\\52\\51\\50
%49\\48\\47\\46\\45\\44\\43\\42\\41\\40
%39\\38\\37\\36\\35\\34\\33\\32\\31\\30
%29\\28\\27\\26\\25\\24\\23\\22\\21\\20
%19\\18\\17\\16\\15\\14\\13\\12\\11\\10
%9\\8\\7\\6\\5\\4\\3\\2\\1\\0

\end{document}